\documentclass[11pt,a4paper,english,superscriptaddress]{revtex4}
\usepackage{amssymb}
\usepackage{amsmath} 
\usepackage{slashed}
\usepackage{graphicx}
\usepackage{babel}
\usepackage{hyperref}
\usepackage{color}
\usepackage{comment}
\makeatletter\AtBeginDocument{\let\@elt\relax}\makeatother

\begin{document}

\title{Gravitational corrections to a non-Abelian gauge theory}

\author{Huan Souza}
\email{huan.souza@icen.ufpa.br}
\affiliation{Faculdade de F\'isica, Universidade Federal do Par\'a, 66075-110, Bel\'em, Par\'a, Brazil.}

\author{L.~Ibiapina~Bevilaqua} \email{leandro.bevilaqua@ufrn.br} \affiliation{Escola de Ci\^encias e Tecnologia, Universidade Federal do Rio Grande do Norte\\ Caixa Postal 1524, 59072-970, Natal, Rio Grande do Norte, Brazil.}

\author{A.~C.~Lehum}
\email{lehum@ufpa.br}
\affiliation{Faculdade de F\'isica, Universidade Federal do Par\'a, 66075-110, Bel\'em, Par\'a, Brazil.}

\begin{abstract}
This paper is part of a series of papers exploring the renormalization of field theories
coupled to gravity using the effective field theory framework. In previous works, we studied
the universality of the electric charge and the two-loops beta function in the Einstein-QED
system. Now, we use this framework to study the non-Abelian gauge theory with fermions
coupled to gravity. We show that even though some of the counterterms are dependent
on the gravitational coupling, the renormalized coupling constant, and therefore the beta
function, do not receive any gravitational correction at one-loop order. Also, we explicitly
show that, at this order, the Slavnov-Taylor identities are respected. Finally, we end the
paper by putting into perspective the results of the present and the past papers.
 
\end{abstract}

\maketitle

\section{Introduction}

Quantum gravity is yet one of the main open questions in high energy physics. The nonrenormalizability of Einstein's theory coupled to other fields \cite{'tHooft:1974bx, PhysRevLett.32.245, Deser:1974cy} was at first considered discouraging for the use of a perturbative treatment for quantum gravity based on general relativity, but our understanding of effective field theories have allowed us to progress, as long as we agree to restrict ourselves to low energies compared to the Planck scale \cite{Donoghue:1994dn, Burgess:2003jk, Shapiro}. With this effective field theory approach, we can study how gravity modifies the renormalization of the coupling constants in a meaningful way.

The renormalization group is a perturbative approach to quantum field theory \cite{Srednicki:2007qs} that consists of two main ideas: to integrate out high-momentum degrees of freedom and a successive rescale of the quantities of the model (coupling constants, wave functions, etc.). By doing so, we end up with renormalized quantities that depend on an arbitrary scale, and thus, we can study how the coupling constants change with this scale, the so-called running of the coupling constants. This has been used in several areas of physics because it is possible to scale the theory as needed to describe a specific system. One example of its use is the description of phase transitions in statistical and condensed matter physics \cite{Goldenfeld, Chaikin}.

We are interested here in the use of the renormalization group to study high-energy physics. In particular, this approach can be used to describe the asymptotic behavior of a theory, for example, how the electrical charge in QED changes when we look at the theory in different energy scales. As it is well known, the coupling constant of a quantum field theory may diverge at some finite energy scale, the so-called Landau pole \cite{Srednicki:2007qs}, but this problem does not occur in asymptotically free theories, such as non-Abelian gauge theory where, as we increase the energy, the coupling gets weaker \cite{Gross:1973id, Politzer:1973fx, Gross:1974jv}.

In 2005, Robinson and Wilczek used the effective field theory prescription of quantum gravity to address the problem of how gravity may influence the asymptotic behavior of gauge theories \cite{Robinson:2005fj}. They suggested that gravitational corrections render all gauge coupling constants asymptotically free. However, this result was soon contested by Pietrykowski \cite{Pietrykowski:2006xy}, who showed that the result was gauge dependent. Subsequently, many works investigate the use of the renormalization group in quantum gravity as an effective field theory~\cite{Felipe:2012vq,Felipe:2013vq,Ebert:2007gf,Nielsen:2012fm,Toms:2008dq,Toms:2010vy,Ellis:2010rw,Anber:2010uj,Bevilaqua:2015hma,Bevilaqua:2021uzk,Bevilaqua:2021uev}.

In a previous work \cite{Bevilaqua:2015hma}, we used dimensional regularization to compute gravitational effects on the beta function of the scalar quantum electrodynamics at one-loop order and found that all gravitational contributions cancel out. At two-loops order, however, we do find nonzero gravitational corrections to the beta function for both scalar and fermionic QED, as shown in a latter work \cite{Bevilaqua:2021uzk}. Despite the gravitational corrections found at two loops, the electrical charge is not asymptotically free neither has a nontrivial fixed point. In fact, those corrections give a positive contribution to the beta function.

Unlike renormalizable field theories, the concept of running coupling may not be useful in the effective field theory approach to quantum gravity because it may be process dependent. The universality of the coupling constants in effective field theories was discussed by Anber \textit{et al.} in \cite{Anber:2010uj}, where it was suggested that an operator mixing would make the coupling constants nonuniversal. This was indeed the case for the quartic self-interaction of scalars in scalar-QED, as discussed in \cite{Bevilaqua:2015hma} but, as shown in \cite{Bevilaqua:2015hma} for scalar-QED and in \cite{Bevilaqua:2021uev} for fermionic-QED it seems not to be the case for the gauge coupling because of Ward identity.

Since gauge symmetry plays an important role in the universality of the gauge coupling for QED, in the present work we will study the gravitational effects on the Slavnov-Taylor identities at one-loop order. The influence of gravity when coupled to a non-Abelian gauge theory was studied in a curved background by Bukhbinder and Odintsov \cite{Buchbinder:1983nug}. The same model was also studied using an expansion around the flat metric in \cite{Tang:2008ah,Tang:2011gz}, where the authors used a method called loop regularization to regulate the divergences. In our work, we add fermions to the non-Abelian gauge theory coupled to gravity and use dimensional regularization to study the Slavnov-Taylor identities and their implications.

The paper is organized as follows. In Sec. \ref{sec2}, we introduce our model and the Slavnov-Taylor identities, which will be important for our analysis. Then, in Sec. \ref{sec3}, we proceed with the renormalization of the model at one-loop order and use the following Sec. \ref{sec3a}, to discuss our results qualitatively. Finally, in Sec. \ref{sec4} we synthesize our results and briefly comment on the literature related to our work. Throughout this work, we use the minimal subtraction (MS) scheme to deal with the divergences, and also, we use natural units $c=\hbar=1$.

\section{The model}\label{sec2}

We start with the Lagrangian describing the effective field theory for a non-Abelian gauge theory with fermions coupled to gravity,
\begin{eqnarray}\label{fQCD}
\mathcal{L}=&& \sqrt{-g}\sum_f\Big\{\frac{2}{\kappa^2}R-\frac{1}{4} g^{\mu\alpha}g^{\nu\beta} G_{\mu\nu}^a G_{\alpha\beta}^a +i\bar{\psi}_f(\nabla_{\mu} - igA_{\mu}^at_a)\gamma^{\mu}\psi_f - m_f\bar{\psi}_f\psi_f +\mathcal{L}_{HO}\Big\}\nonumber\\
&&~
\end{eqnarray}
\noindent where the index $f=1,2,\cdots,N_f$ runs over the fermions flavors and $G^a_{\mu\nu}=\nabla_\mu A_\nu^a-\nabla_\nu A_\mu^a + gf^{abc}A^b_\mu A^c_\nu$ is the non-Abelian field-strength with $f^{abc}$ being the structure constants of the $SU(N)$ group. $\mathcal{L}_{HO}$ is the Lagrangian of higher order terms, and the relevant terms for our purposes are explicitly written bellow,
\begin{eqnarray}\label{ho}
\mathcal{L}_{HO} & = & i\bar\psi_f~\frac{\Box}{M_P^2}\left(\tilde{g}_1\slashed{\partial} - \tilde{g}_2 m_f\right)\psi_f-\frac{\tilde{g}_3}{4M_P^2}G^{\mu\nu}_a\Box G_{\mu\nu}^a \nonumber \\
& & \qquad\qquad\qquad\qquad\qquad + \frac{i\tilde{g}_4}{2M_P^2}\bar\psi_f\gamma_\mu\partial_\nu\psi_f t^aG^{\mu\nu}_a+ \frac{\tilde{g}_5}{M_P^2}(\bar{\psi}_f\gamma^\mu\psi_f)^2 + \cdots
\end{eqnarray}

In the above expression for $\mathcal{L}_{HO}$, $\tilde{g}_i$ are dimensionless coupling constants, $M_P$ is the Planck mass, and the spacetime (greek) indices are raised and lowered with the flat metric $\eta_{\mu\nu}=(+,-,-,-)$. Now we expand $g_{\mu\nu}$ around the flat metric as
\begin{equation}\label{metric}
 g_{\mu\nu} = \eta_{\mu\nu} + \kappa h_{\mu\nu} \quad (\text{exactly}), \qquad g^{\mu\nu} = \eta^{\mu\nu} - \kappa h^{\mu\nu} +\cdots
\end{equation}
such that
\begin{equation}\label{sqrt}
 \sqrt{-g} = 1 + \frac{\kappa}{2}h + \cdots,
\end{equation}
\noindent where $h = \eta^{\mu\nu}h_{\mu\nu}$. Also, the affine connection is written as
\begin{equation}\label{connection}
 \Gamma^\lambda_{~\mu\nu} = \frac{1}{2}\kappa(\eta^{\lambda\sigma} - \kappa h^{\lambda\sigma})(\partial_\mu h_{\sigma\nu} + \partial_\nu h_{\sigma\mu} - \partial_\sigma h_{\mu\nu}).
\end{equation}

Here, we will follow the results in Ref. \cite{Choi:1994ax}. Then, we can organize the Lagrangian as follows (before quantization),
\begin{subequations}
 \begin{eqnarray}
  \mathcal{L} &=& \mathcal{L}_h + \mathcal{L}_f + \mathcal{L}_A;\\
  \mathcal{L}_h &=& \frac{2}{\kappa^2}\sqrt{-g}R;\\
  \mathcal{L}_f &=& \sqrt{-g}[i\bar{\psi}_f(\nabla_\mu - igA_\mu^at^a)\gamma^\mu\psi_f - m_f\bar{\psi}_f\psi_f];\\
  \mathcal{L}_A &=& -\frac{\sqrt{-g}}{4}g^{\mu\alpha}g^{\nu\beta}G_{\mu\nu}^aG^a_{\alpha\beta}.
 \end{eqnarray}
\end{subequations}

Then, using Eqs. \eqref{metric}, \eqref{sqrt}, and \eqref{connection}, we obtain for the gravity sector
\begin{subequations}\label{Lh}
 \begin{eqnarray}
  \mathcal{L}_h &=& \mathcal{L}_h^0 + \kappa\mathcal{L}_h^1 + \cdots\\
  \mathcal{L}_h^0 &=& -\frac{1}{4}\partial_\mu h\partial^\mu h + \frac{1}{2}\partial_\mu h^{\sigma\nu}\partial^{\mu}h_{\sigma\nu};\\
  \mathcal{L}_h^1 &=& \frac{1}{2}h^\alpha_{~\beta}\partial^\mu h^\beta_{~\alpha}\partial_\mu h - \frac{1}{2}h^\alpha_{~\beta}\partial_\alpha h^\mu_{~\nu}\partial^\beta h^\nu_{~\mu} - h^\alpha_{~\beta}\partial_\mu h^\nu_{~\alpha}\partial^\mu h^\beta_{~\nu}\nonumber\\
  && +\frac{1}{4}h\partial^\beta h^\mu_{~\nu}\partial_\beta h^\nu_{~\mu} + h^\beta_{~\mu}\partial_\nu h^\alpha_{~\beta}\partial^\mu h^\nu_{~\alpha} - \frac{1}{8}h\partial^\nu h\partial_\nu h
 \end{eqnarray}
\end{subequations}
for the fermionic sector,
\begin{subequations}
 \begin{eqnarray}
  \mathcal{L}_f &=& \mathcal{L}_f^0 + g\bar{\psi}_f\gamma^\mu\psi_f A_\mu^a t^a + \kappa\mathcal{L}_f^1 + \cdots\\
  \mathcal{L}_f^0 &=& \frac{i}{2}(\bar{\psi}_f\gamma^\mu\partial_\mu\psi_f - \partial_\mu\bar{\psi}_f\gamma^\mu\psi_f) - m_f\bar{\psi}_f\psi_f;\\
  \mathcal{L}_f^1 &=& \frac{1}{2}h\mathcal{L}_f^0 - \frac{i}{4}h_{\mu\nu} (\bar{\psi}_f\gamma^\mu\partial^\nu\psi_f- \partial^\nu\bar{\psi}_f\gamma^\mu\psi)
 \end{eqnarray}
\end{subequations}
and finally, for the gauge sector,
\begin{subequations}\label{LA}
 \begin{eqnarray}
  \mathcal{L}_A &=& \mathcal{L}_A^0 + \kappa\mathcal{L}_A^1 + \cdots\\
  \mathcal{L}_A^0 &=& -\frac{1}{4}G_{\mu\nu}^aG^{\mu\nu}_a \\
  \mathcal{L}_A^1 &=& \frac{1}{2}h^\tau_{~\nu}G^{\mu\nu}_aG_{\mu\tau}^a + \frac{1}{2}h\mathcal{L}_A^0.
 \end{eqnarray}
\end{subequations}

Now, we proceed to the quantization of the model following the Faddeev-Popov procedure, in which we must introduce the gauge-fixing and the ghosts, both for the vector and the tensor fields. Since we are working with the one-graviton exchange approximation, these terms will not contribute to the renormalization of the gauge coupling constant. In fact, the introduction of ghosts for the tensor fields leads to interactions like $\bar{c}_hc_hh$ (with $c_h$ being the graviton's ghost), so we have an interaction of two ghosts and one graviton. Therefore, we will not write them explicitly and the ghosts Lagrangian will be written only for the gauge sector, that is the usual one for non-Abelian gauge theories,
\begin{equation}\label{ghost}
 \mathcal{L}_{ghost} = \sqrt{-g}\Bigr\{-g_{\mu\nu}\partial^{\mu}\bar{c}^a\partial^{\nu}c^a + gg_{\mu\nu}f^{abc}A^{\mu}_a\partial^{\nu}\bar{c}_b c_c\Bigr\}.
\end{equation}

Using the expansion around the flat metric we end up with
\begin{subequations}
 \begin{eqnarray}
  &&\mathcal{L}_{ghost} = \mathcal{L}_g^0 + \kappa\mathcal{L}_g^1;\\
  &&\mathcal{L}_{g}^0=-\partial_\mu \bar{c}^a\partial^\mu c^a + gf^{abc}A^\mu_a(\partial_\mu\bar{c}_b) c_c;\\
  &&\mathcal{L}_g^1 = \frac{1}{2}h\mathcal{L}_g^0 + h_{\mu\nu}(-\partial^\mu\bar{c}^a\partial^\nu c^a + gf^{abc}A^{\mu}_a(\partial^\nu \bar{c}_b)c_c).
 \end{eqnarray}
\end{subequations}

The propagators for leptons, gluons, ghosts, and graviton are given, respectively, by
\begin{subequations}\label{propagators}
\begin{eqnarray}
S_F(p) &=& i\frac{\slashed{p}+m_l}{p^2-m_l^2};\\
\Delta^{\mu\nu}_{ab}(p) &=& \frac{i}{p^2}\left(\eta^{\mu\nu}-(1-\xi_A)\frac{p^\mu p^\nu}{p^2} \right)\delta_{ab};\\
\Delta_{ab}(p) &=& \frac{i}{p^2}\delta_{ab};\\ 
\Delta^{\alpha\beta\mu\nu}(p) &=& \frac{i}{p^2}\left(P^{\alpha\beta\mu\nu}-(1-\xi_h)\frac{Q^{\alpha\beta\mu\nu}}{p^2}\right).  
\end{eqnarray}
\end{subequations}
\noindent Notice that we have not chosen any specific gauge for the gluon and graviton propagators, $\xi_A$ and $\xi_h$ being their gauge fixing parameters. The projectors $P^{\alpha\beta\mu\nu}$ and $Q^{\alpha\beta\mu\nu} $ are given by
\begin{eqnarray}
P^{\alpha\beta\mu\nu} &=&\frac{1}{2} \left(\eta^{\alpha\mu}\eta^{\beta\nu}+\eta^{\alpha\nu}\eta^{\beta\mu}-\eta^{\alpha\beta}\eta^{\mu\nu} \right);\nonumber\\
Q^{\alpha\beta\mu\nu} &=& (\eta^{\alpha\mu}p^\beta p^\nu+\eta^{\alpha\nu}p^\beta p^\mu+\eta^{\beta\mu}p^\alpha p^\nu+\eta^{\beta\nu}p^\alpha p^\mu).
\end{eqnarray}

In order to study the renormalization of the model, we redefine the fields and parameters in the Lagrangian \eqref{fQCD}. For example, the vector, fermion and scalar field strengths are redefined as $A^\mu_a\rightarrow Z_3^{1/2}A^\mu_a$ and $\psi_f\rightarrow Z_{2f}^{1/2}\psi_f$, where $Z$ are the renormalizing functions, organized as a perturbative series,
\begin{eqnarray}
Z=Z^{(0)}+ Z^{(1)}+Z^{(2)}+\cdots, \qquad \text{with} \quad Z^{(0)} = 1.
\end{eqnarray}
The relation between the bare ($g_0$) and the renormalized ($g$) gauge coupling constants in terms of the $Z$ functions may be written in four different ways. We then have
\begin{subequations}\label{g0}
\begin{eqnarray}\label{eq_e_0}
g&=&\mu^{-2\epsilon}\frac{Z_2 Z_3^{1/2}}{Z_1}g_0;\\
g&=&\mu^{-2\epsilon}\frac{Z_{3}^{3/2}}{Z_{3g}}g_0;\label{g0Z3g}\\
g&=&\mu^{-2\epsilon}\frac{Z_{3}}{Z_{4g}^{1/2}}g_0;\label{g0Z4}\\
g&=&\mu^{-2\epsilon}\frac{Z_{2c}Z_3^{1/2}}{Z_{1c}}g_0,
\end{eqnarray}
\end{subequations}
\noindent where $\mu$ is a mass scale introduced by the dimensional regularization (DR), used to regularize the UV divergences in the Feynman amplitudes, $\epsilon$ is related to the spacetime dimension D by D$=4-2\epsilon$, $Z_1$ is the gauge coupling constant counterterm, $Z_{3g}$ is the counterterm that renormalized the three-point function of the gluons, $Z_{4g}$ is the counterterm that renormalizes the four-point functions of the gluons, $Z_{2c}$ is the wave function counterterm for the ghosts and $Z_{1c}$ is the counterterm that renormalizes the gluon-ghost vertex. 

Together, Eqs. \eqref{g0} provide relations between the Green functions that must be satisfied to ensure gauge invariance, the so-called Slavnov-Taylor identities \cite{Slavnov:1972fg, Taylor:1971ff}. In terms of the renormalizing functions $Z$, the relations are
\begin{eqnarray}\label{Slavnov-Taylor}
 \frac{Z_1}{Z_2} &=& \frac{Z_{3g}}{Z_3}~ = \frac{Z_{4g}^{1/2}}{Z_3^{1/2}} = \frac{Z_{1c}}{Z_{2c}}.
\end{eqnarray}

It is important to notice that the above relations are tied to the gauge invariance of the theory and we expect that they will be true at any order in perturbation theory. In the next section we will show that they are indeed respected at one-loop.

\section{The one-loop renormalization}\label{sec3}

Our goal in this section is to show explicitly that the Slavnov-Taylor identities are respected at one-loop order. In order to do that, we will compute the fermion, vector, and ghost fields self-energies ($\Sigma_f, \Pi^{\mu\nu}_{ab}$ and $\Sigma_{ab}$, respectively), also the quark-gluon, ghost-gluon and gluon-gluon three-point functions ($\Gamma^{\mu}_a$, $\Gamma^\mu_{abc}$ and $\Pi^{\mu\nu\alpha}_{abc}$, respectively), and finally, the gluon four-point function ($\Gamma^{\mu\nu\rho\sigma}_{abcd}$). All the computations were done using the \textit{Mathematica} packages: \textit{FeynRules} to generate the models \cite{feynrules}, \textit{FeynArts} to draw the diagrams \cite{Hahn:2000kx}, and \textit{FeynCalc} to simplify and compute the amplitudes \cite{Shtabovenko:2020gxv}. The files we used can be found in \cite{site_lehum}.

\begin{figure}[t]
	\begin{center}
	\includegraphics[angle=0 ,width=10cm]{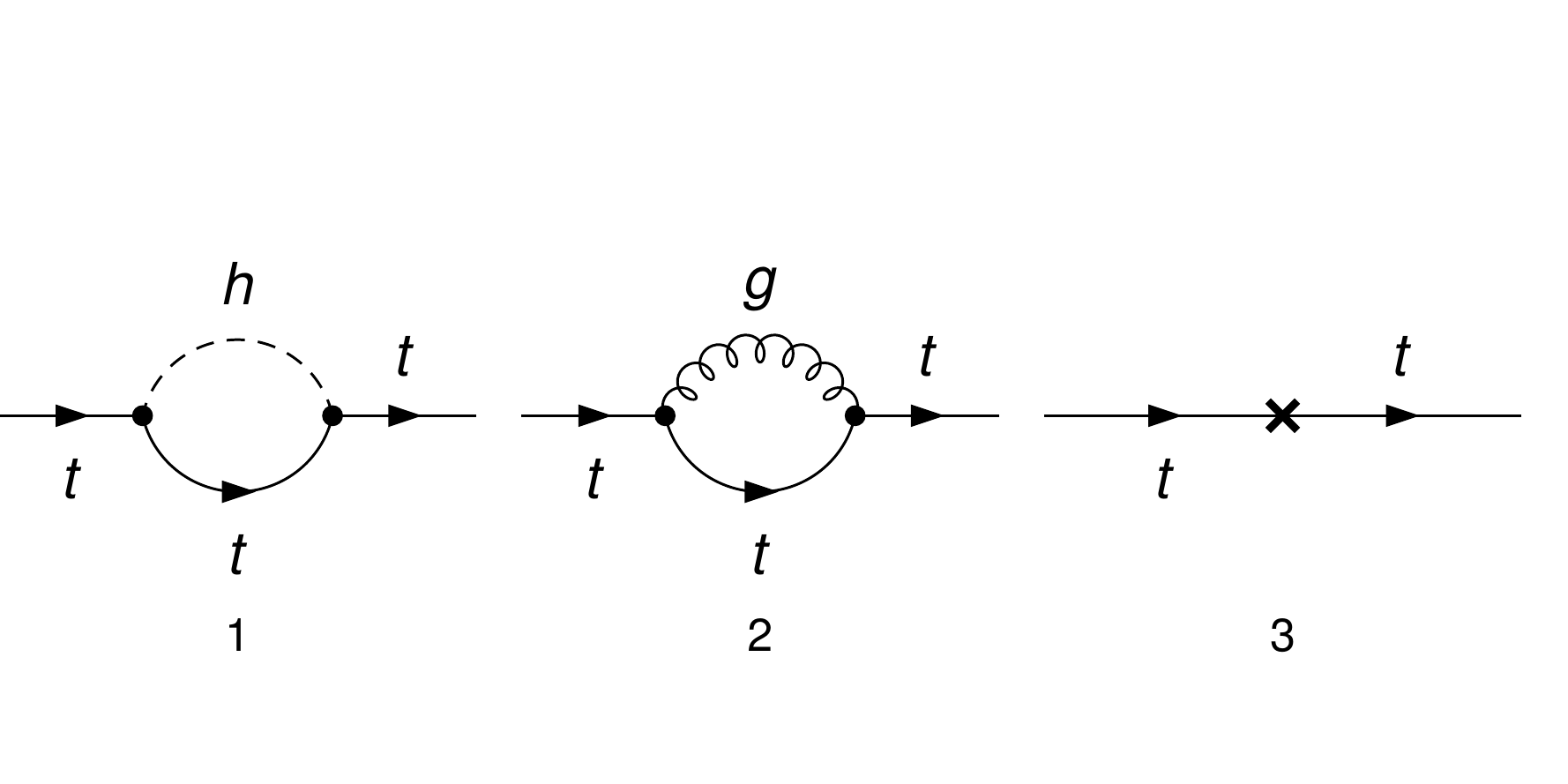}
	\caption{Feynman diagrams for the quark top self-energy. Continuous, wiggly, dotted, and dashed lines represent the fermion, gluon, ghost, and graviton propagators, respectively. The $t$ label represents the quark-top, since this is the most massive quark and therefore, will have the major gravitational contributions. For the others self-energies, we just have to change the quark-top propagators for the other ones.}	\label{fig01}
	\end{center}
\end{figure}

Thus, let us first calculate at one-loop order the relevant $Z$ factors for the renormalization of the fermionic model, starting with the one-loop corrections to the self-energy diagrams. The expression corresponding to the self-energies of the quarks, Fig.\ref{fig01}, can be cast as
\begin{eqnarray}\label{1eq02}
-i\Sigma_f(p) &=& \frac{i \left(C_A-2 C_F\right) \slashed{p} \left(16 \left(C_A^2-1\right) g_s^2 \xi_A+37 \kappa ^2 C_A m_f^2-29\kappa ^2 C_A \xi_{h} m_f^2\right)}{512 \pi ^2 \epsilon }\nonumber\\
&&-\frac{i m_t \left(C_A-2 C_F\right) \left(16 \left(C_A^2-1\right) g_s^2 \left(\xi _A+3\right)+46\kappa ^2 C_A m_f^2-38\kappa ^2 C_A \xi _h m_f^2\right)}{512 \pi ^2 \epsilon }\nonumber\\
&&-i C_A Z_{m_f}^{(1)} m_t \left(C_A-2 C_F\right) +i C_A Z_{2f}^{(1)} \left(C_A-2 C_F\right) \slashed{p} + HO + \mathrm{finite},
\end{eqnarray}
\noindent where $C_A=N$ for the $SU(N)$ group and $C_F=\frac{C_A^2-1}{2C_A}$. The terms proportional to $\slashed{p}$ are renormalized, through MS, by the counterterm $Z_{2f}^{(1)}$, the terms proportional to $\slashed{p}^0$ are renormalized by the counterterm $Z_{m_f}^{(1)}$ and $HO$ represents the higher order terms in $p$ that are renormalized by the high derivative operators, Eq. (\ref{ho}). 

Imposing finiteness to $\Sigma_f(p)$, we find the following one-loop counterterms 
\begin{subequations}\label{ct01}
\begin{eqnarray}
Z_{2f}^{(1)} &=& \frac{\kappa ^2 m_f^2 \left(29 \xi _h-37\right)}{512 \pi ^2 \epsilon }-\frac{C_F g_s^2 \xi _A}{16 \pi ^2 \epsilon },\label{eq_z2f}\\
Z_{m_f}^{(1)} &=& \frac{\kappa ^2 m_f^2 \left(19 \xi _h-23\right)}{256 \pi ^2 \epsilon }-\frac{C_F g_s^2 \left(\xi _A+3\right)}{16 \pi ^2 \epsilon }.\label{ct01-2}
\end{eqnarray}
\end{subequations}

\begin{figure}[t!]
	\begin{center}
	\includegraphics[angle=0 ,width=10cm]{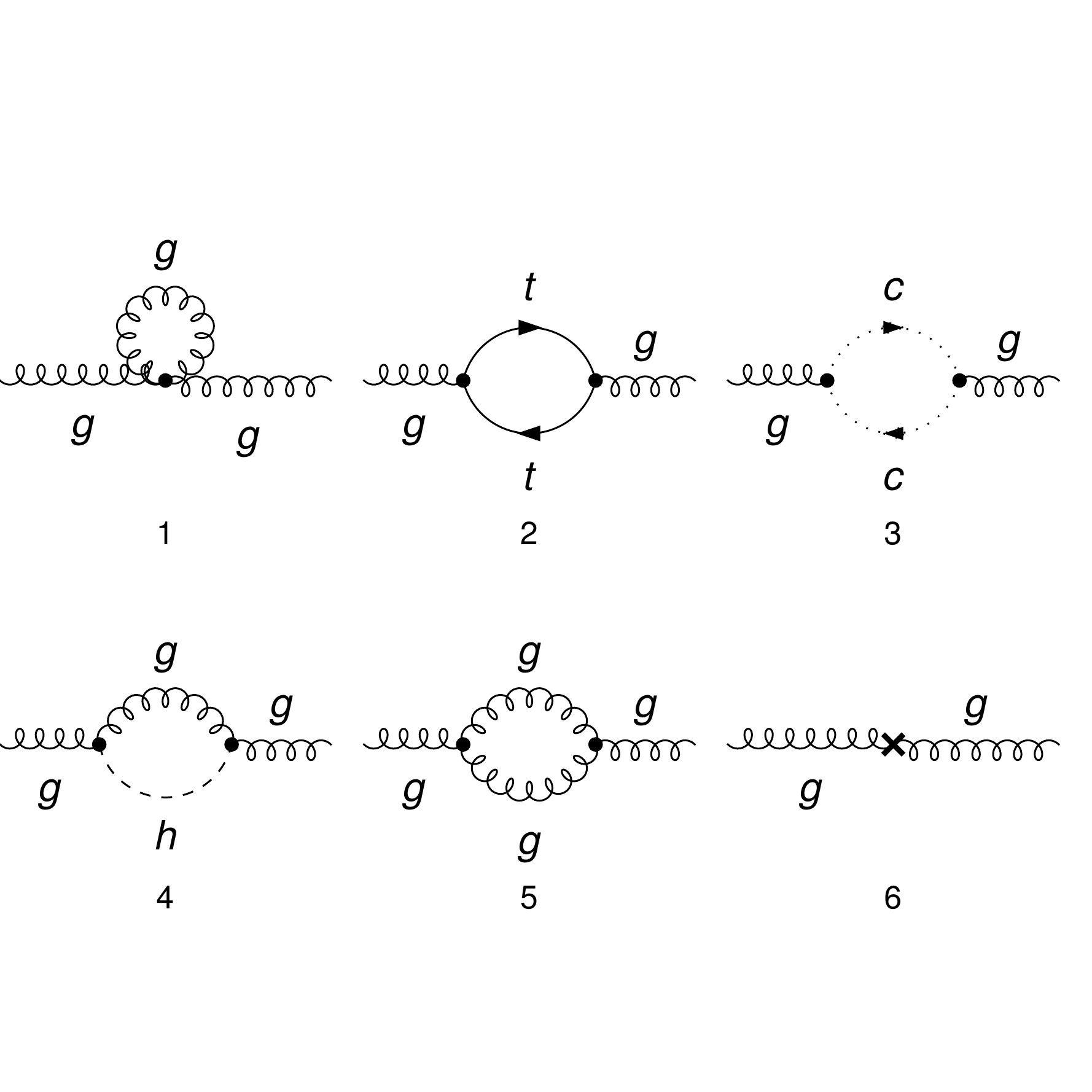}
	\caption{Feynman diagrams for the gluon self-energy.}
	\label{fig02}
	\end{center}
\end{figure}

For the gluon self-energy, we write the one-loop correction (corresponding to the diagrams in Fig.\ref{fig02}) as
\begin{eqnarray}\label{1eq02b}
\Pi^{\mu\nu}_{ab}(p)= \left(p^2 \eta^{\mu  \nu }-p^{\mu } p^{\nu }\right)\Pi(p)\delta_{ab},
\end{eqnarray}
\noindent where
\begin{eqnarray}\label{eq_pi1}
\Pi(p)= -Z_3^{(1)}-\tilde{Z}_3^{(1)}p^2-\frac{\left(3 C_A g_s^2 \xi _{A}-13 C_A g_s^2+4 N_f g_s^2-2 \kappa ^2 p^2+3 \kappa ^2 p^2 \xi _{h}\right)}{96 \pi ^2 \epsilon }+\mathrm{finite},
\end{eqnarray}
from which we can see that $Z_3$ is the renormalizing factor for the quadratic term $G^{\mu\nu}_a G_{\mu\nu}^a$, while $\tilde{Z}_3$ renormalizes a higher derivative term like $G^{\mu\nu}_a\Box G_{\mu\nu}^a$. Thus, $Z_3$ is the relevant counterterm to the beta function of the color charge. Notice that the UV divergent part of Eq.(\ref{eq_pi1}) is not dependent on the masses of the quarks.

\begin{figure}[h!]
	\begin{center}
	\includegraphics[angle=0 ,width=12cm]{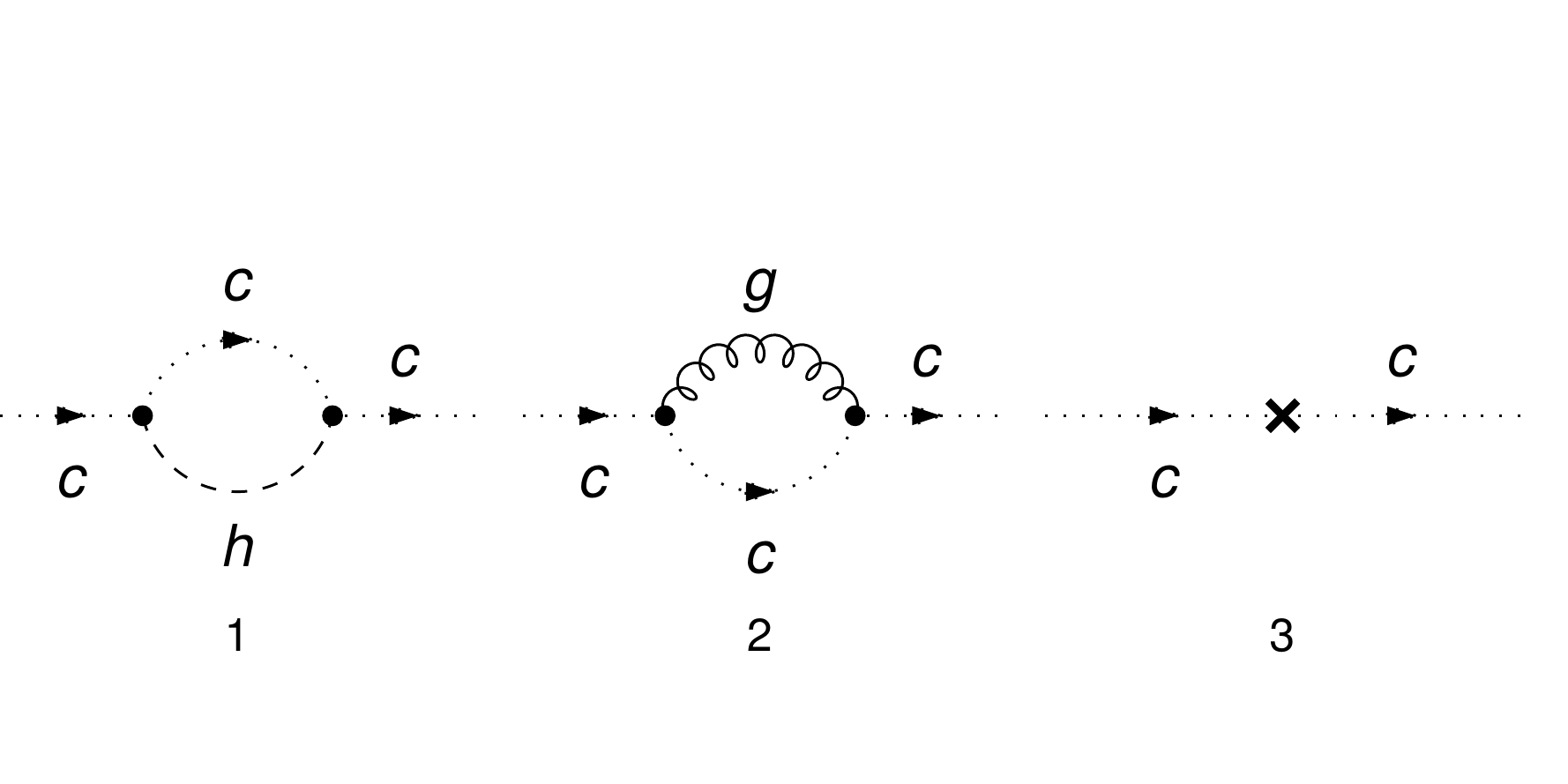}
	\caption{Feynman diagrams for the ghost self-energy.}
	\label{fig03}
	\end{center}
\end{figure}

Imposing the finiteness over $\Pi(p)$, we find 
\begin{subequations}\label{ct02}
\begin{eqnarray}
Z_3^{(1)} &=& -\frac{g_s^2 \left(3 C_A \xi _{A}-13 C_A+4 N_f\right)}{96 \pi ^2 \epsilon },\label{eq_z3}\\
\tilde{Z}_3^{(1)} &=& -\frac{\kappa^2(3\xi_h-2)}{96\pi^2\epsilon}~.\label{ct02-2}
\end{eqnarray}
\end{subequations}

Contributions to the ghost self-energy up to one-loop order are depicted in Fig. \ref{fig03}. The resulting expression is
\begin{eqnarray}\label{ghostSE}
-i\Sigma_{ab} &=& \frac{i p^2 C_A g_s^2 \delta_{ab} \xi _{A}}{64 \pi ^2 \epsilon }-\frac{3 i p^2 C_A g_s^2 \delta_{ab}}{64 \pi ^2 \epsilon }+i p^2 Z_{2c}^{(1)} \delta_{ab}-\frac{3 i \kappa ^2 p^4 \delta _{ab}}{32 \pi ^2 \epsilon }-\frac{i \kappa ^2 p^4 \xi _{h} \delta_{ab}}{8 \pi ^2 \epsilon } +\mathrm{finite},
\end{eqnarray}
and, imposing finiteness, we find
\begin{eqnarray}\label{eq_z2c}
Z_{2_c}^{(1)} &=& -\frac{C_A g_s^2 \left(\xi _A-3\right)}{64 \pi ^2 \epsilon }.
\end{eqnarray}

The gravitational contributions will, therefore, be renormalized by a higher-order term. This is expected since both the ghosts and the graviton are massless. Because of that, the only contribution proportional to $\kappa^2$ must be of the order $p^4$.
\begin{figure}[t!]
	\begin{center}
	\includegraphics[angle=0 ,width=10cm]{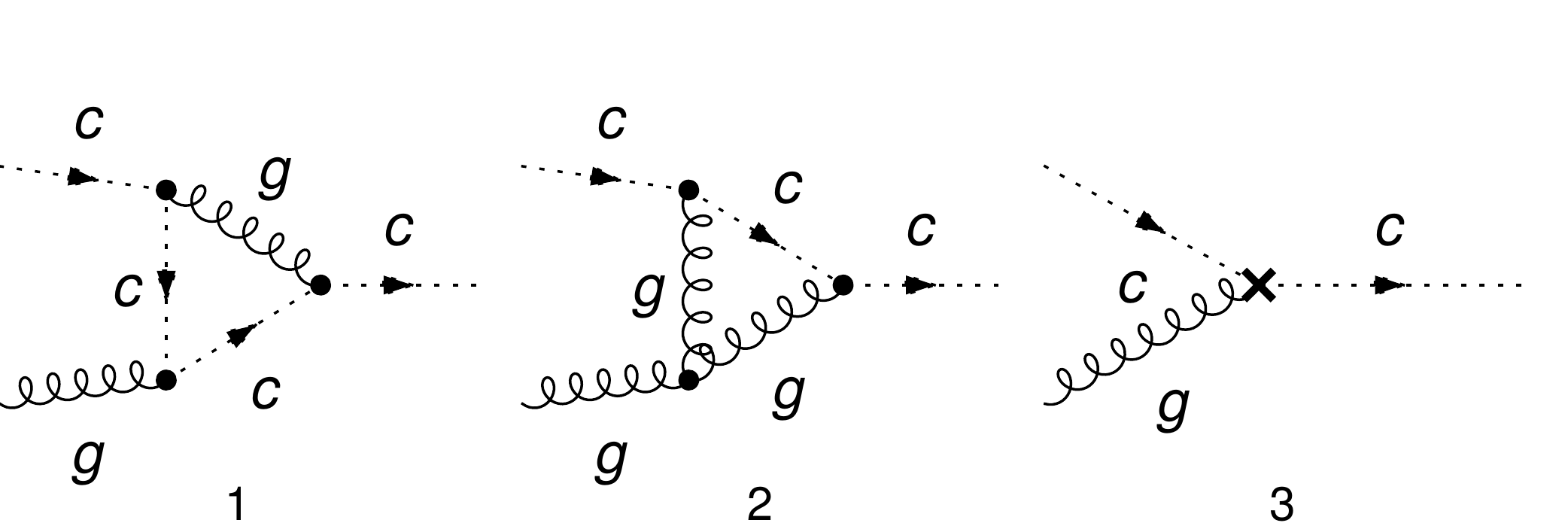}
	\caption{Feynman diagrams for the vertex interaction between gluons and ghosts up to one-loop order.}
	\label{fig04}
	\end{center}
\end{figure}

Now we proceed to the vertices diagrams, starting with the contributions to the ghost-gluon vertex (Fig. \ref{fig04}). From now on, $p_1$ and $p_2$ represent incoming external momenta, and $p_3$ and $p_4$ outgoing momenta. The result for these diagrams is
\begin{equation}
 \Gamma_{abc}^\mu = -g p_3^{\mu } f_{abc}\left(\frac{C_A g^2\xi_A}{32 \pi ^2 \epsilon } + Z_{1c}^{(1)}\right) + O(p^3) + \text{finite}.
\end{equation}
\noindent Then, subtracting the UV pole, we have 
\begin{equation}\label{eq_z1c}
 Z_{1_c}^{(1)} = -\frac{C_A g^2\xi_A}{32 \pi ^2 \epsilon }.
\end{equation}

Once again, all the gravitational corrections are renormalized by higher-order terms.
\begin{figure}[t!]
	\begin{center}
	\includegraphics[angle=0 ,width=10cm]{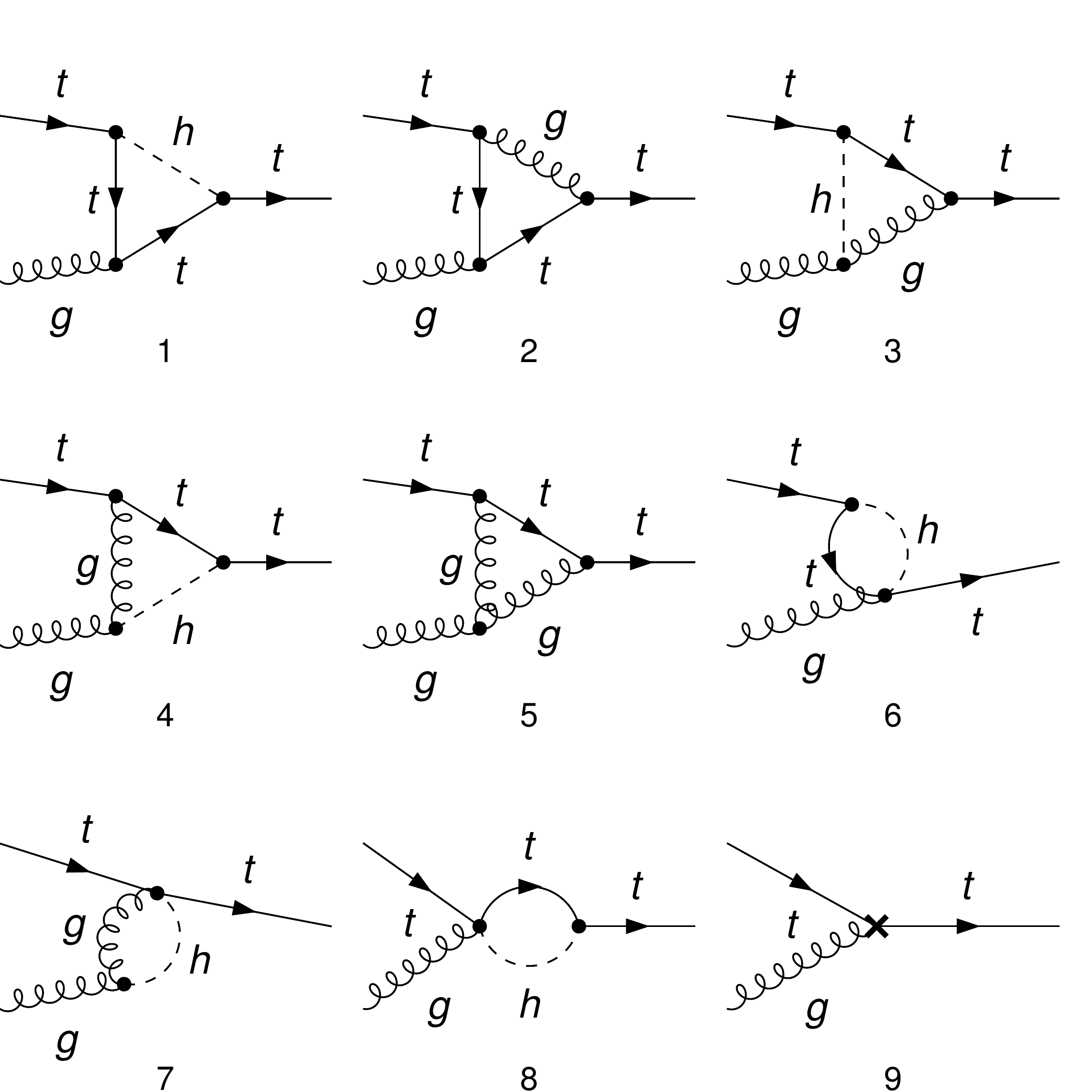}
	\caption{Feynman diagrams to the vertex interaction between quarks top and gluons up to one-loop order.}
	\label{fig05}
	\end{center}
\end{figure}

Contributions to the quark-gluon vertex function up to one-loop order are depicted in Fig. \ref{fig05}, and the resulting expression is
\begin{eqnarray}
 -i\Gamma^\mu_a&=&-i \gamma ^{\mu } g t_{a} \left(C_A-2 C_F\right)\left[\frac{8\left(\left(3 C_A^2-2\right)\xi_A + 3C_A^2\right) g^2+\kappa ^2 C_A m^2(37-29\xi_h)}{512 \pi ^2 \epsilon }+Z_1^{(1)} C_A\right]\nonumber\\
 &&+ O(p) + \text{finite},
\end{eqnarray}

\noindent from which, through MS, we find
\begin{equation}\label{eq_z1}
 Z_1^{(1)}=\frac{\kappa ^2 m_f^2 \left(29 \xi _h-37\right)-8 g_s^2 \left(\xi _A \left(C_A+4 C_F\right)+3 C_A\right)}{512 \pi ^2 \epsilon }.
\end{equation}

For the gluon vertex (Fig. \ref{fig06} showed at the end of the paper), we have used the projection
\begin{equation}
 \Pi^{\mu\nu\alpha}_{abc}=\eta^{\mu\nu}\Pi^{\alpha}_{abc} \qquad \Rightarrow \qquad \Pi^\alpha_{abc} = \frac{1}{4}\eta_{\mu\nu}\Pi^{\mu\nu\alpha}_{abc}
\end{equation}
\noindent to simplify our computations. Using the fact that $p_3 = p_1 + p_2$, we find
\begin{eqnarray}
 -i\Pi^\mu_{abc} &=& -\frac{g_s^3 \left(p_1-p_2\right){}^{\alpha } f^{abc} \left(9 C_A \xi _{A}-17 C_A+8 N_f\right)}{256 \pi ^2 \epsilon }-\frac{3}{4} Z_{3g}^{(1)} g_s \left(p_1-p_2\right){}^{\alpha } f^{abc}\nonumber\\
 &&+O(p^2)+\mathrm{finite},
\end{eqnarray}
\noindent and, imposing finiteness through MS, we have
\begin{equation}\label{eq_z13g}
 Z_{3g}^{(1)} = -\frac{g_s^2 \left(9 C_A \xi _A-17 C_A+8 N_f\right)}{192 \pi ^2 \epsilon }.
\end{equation}

Finally, we consider the scattering of gluons (Fig. \ref{fig07} showed at the end of the paper) and compute the gluon four-point function counterterm. Since the interaction of four gluons has no derivative, $Z_{4g}$ will renormalize terms proportional to $p^0$. Therefore, we can set external momentum equals to zero if we restrict ourselves to the computation of this counterterm. Also, for simplicity, we have used the scalar projection
\begin{equation}\label{p4g}
 \Gamma_{abcd} = \frac{1}{16}\eta_{\mu\nu}\eta_{\rho\sigma}\Gamma^{\mu\nu\rho\sigma}_{abcd},
\end{equation}
\noindent to obtain
\begin{eqnarray}
 -i\Gamma_{abcd} &=& -\left(\frac{i g_s^4 \left(3 C_A \xi _A-2 C_A+2 N_f\right)}{32 \pi ^2 \epsilon }+\frac{3}{2} i Z_{4g}^{(1)} g_s^2\right)\Bigr( \text{tr} (t_{a}t_{b}t_{c}t_{d})-2 \text{tr} (t_{a}t_{c}t_{b}t_{d})-2 \text{tr}(t_{b}t_{c}t_{a}t_{d})\nonumber\\
 &&+ \text{tr}(t_{b}t_{a}t_{c}t_{d}) + \text{tr} (t_{c}t_{a}t_{b}t_{d})+\text{tr} (t_{c}t_{b}t_{a}t_{d})\Bigr)
\end{eqnarray}
\noindent Then, imposing finiteness through MS, we have
\begin{equation}\label{eq_z14g}
 Z^{(1)}_{1_{4g}}=-\frac{g_s^2 \left(3 C_A \xi _A-2 C_A+2 N_f\right)}{48 \pi ^2 \epsilon }.
\end{equation}

As we can see from Eqs. \eqref{eq_z2f},\eqref{eq_z3},\eqref{eq_z2c},\eqref{eq_z1c},\eqref{eq_z1},\eqref{eq_z13g},\eqref{eq_z14g}, the Slavnov-Taylor identities are respected, since we have
\begin{equation}
 Z_1^{(1)}-Z_2^{(1)} = Z_{3g}^{(1)} - Z_3^{(1)} = \frac{1}{2}\left(Z_{4g}^{(1)} - Z_3^{(1)}\right) = Z_{1c}^{(1)} - Z_{2c}^{(1)} = -\frac{C_A g_s^2(3+\xi_A)}{64 \pi^2 \epsilon },
\end{equation}
\noindent indicating that gravitational interaction does not spoil the gauge symmetry and we can define a global color charge.

Moreover, we can show that the beta function is independent of $\kappa$ and $m_f$, as the expression the one-loop beta function of the color charge can be found through Eq. \eqref{eq_e_0} and can be cast as
\begin{eqnarray}
 \beta(g) &=& \lim_{\epsilon\rightarrow0}\mu\frac{dg}{d\mu}=\lim_{\epsilon\rightarrow0}\mu\frac{d}{d\mu}\left[g_{0}\left(1-Z_1^{(1)}+Z_2^{(1)}+\frac{Z_3^{(1)}}{2}\right)\mu^{-2\epsilon}\right]\nonumber\\
 &=& -\frac{g^3}{(4\pi)^2}\left(\frac{11}{3}C_A-\frac{4}{6}N_f\right).
\end{eqnarray}

This result is also gauge independent, just as we have found at one-loop order for the scalar and the fermionic QED~\cite{Bevilaqua:2015hma,Bevilaqua:2021uev}. 

The authors in Ref.\cite{Folkerts:2011jz} showed that in the weak-gravity limit there is no gravitational contribution at one-loop order if the regularization scheme preserves the symmetries of the model, such as dimensional regularization. On the other hand, if the regularization scheme does not preserve all the symmetries it gives a negative contribution to the beta function (as done in \cite{Robinson:2005fj}). At two-loops, based on our results for the abelian case \cite{Bevilaqua:2021uzk} and the considerations developed in the next section, we expect that, in the non-Abelian case, the gravitational contribution to the beta function should be positive but too small to spoil the asymptotic freedom of the theory.

\section{Qualitative view of the gravitational corrections}\label{sec3a}

In this section, we discuss a qualitative view of quantum corrections to get a better understanding of why there are no gravitational contributions to the renormalized gauge coupling constant at one-loop order while we expect that there will be positive contributions to the beta function at two-loops. For simplicity, we will consider the QED case worked out in \cite{Bevilaqua:2021uzk} and then extend our conclusions to a non-Abelian theory.

\begin{figure}[t!]
	\begin{center}
	\includegraphics[angle=0 ,width=6cm]{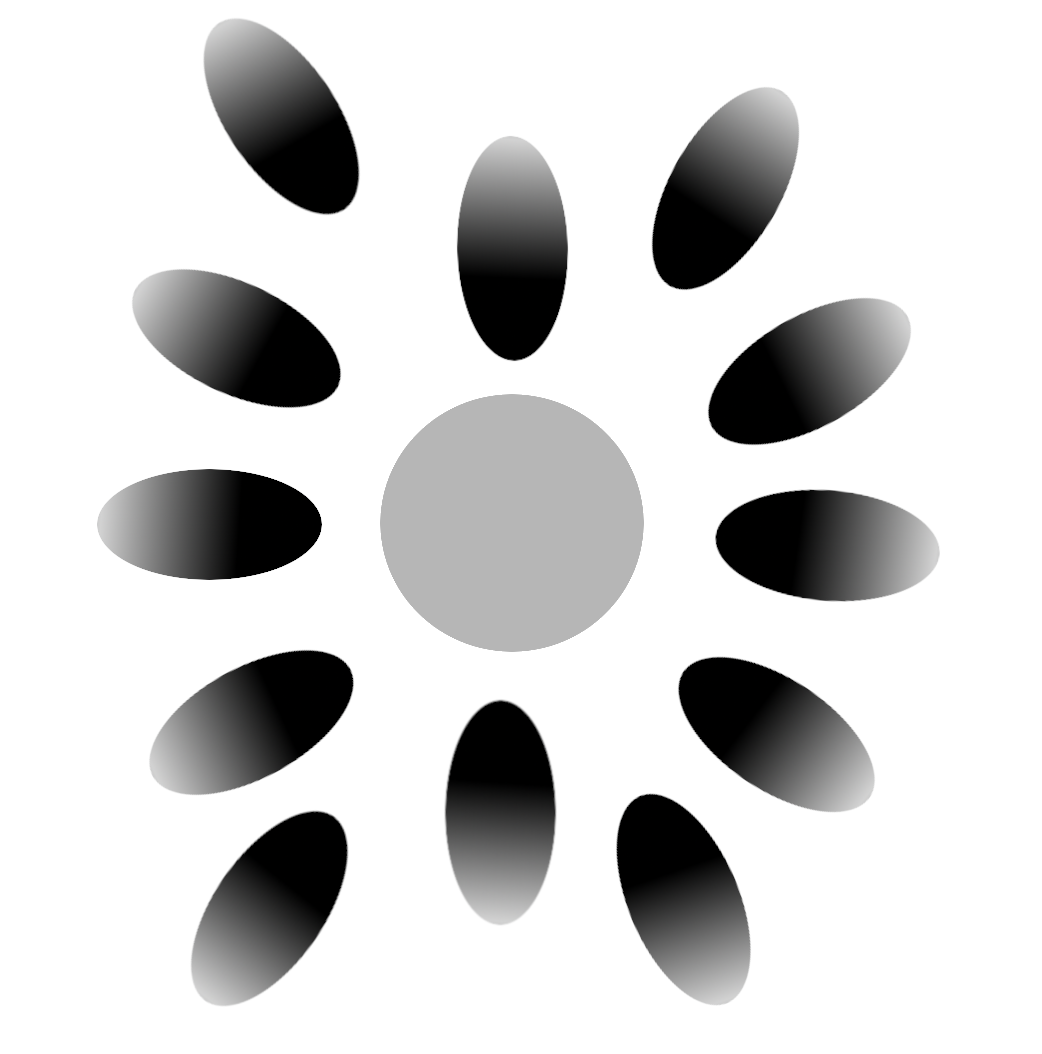}
	\caption{The gray color represents negative charges while the black color represents positive charges. The central circle is an electron surrounded by polarized virtual particles.}
	\label{polarization}
	\end{center}
\end{figure}

In QED, we can think about the renormalization of the electric charge as the polarization of the virtual particles in the vacuum, resulting in a weaker coupling (see Fig. \ref{polarization}). This quantum phenomenon is analogous to the polarization that takes place in the study of classical electrodynamics in ponderable media. In fact, when a charge is placed in a dielectric medium, it will polarize the charges of the material and this polarization will make the macroscopic electric field weaker than it would be in the vacuum. 

Let us consider linear dielectrics and write the permittivity as
\begin{equation}
 \epsilon = \epsilon_0(1+\eta_e),
\end{equation}
where $\epsilon_0$ is the electrical permittivity of the vacuum and $\eta_e$ is the electric susceptibility. This quantity indicates the degree of polarization of a material in response to an applied electric field and depends both on the internal structure of the material (how the charges are distributed) and external properties (such as temperature). 

The effective electrical field inside the material is given by
\begin{equation}\label{eff-electric}
 \textbf{E} = \frac{1}{\epsilon_r}\textbf{E}_{vacuum}
\end{equation}
where $\epsilon_r \equiv \epsilon/\epsilon_0$ is the relative electrical permittivity. Since $\eta_e>0$ and $\epsilon_r>1$, the effective electrical field $\textbf{E}$ is weaker then $\textbf{E}_{vacuum}$. Moreover, if there are more charges to be polarized (an electrically denser material), we will have a bigger electric susceptibility and thus a weaker effective field.

In quantum field theory, we can think about the vacuum itself as a medium to be polarized (but this time, the polarization is of virtual particles) and the effective field \eqref{eff-electric} as analogous to the renormalized field. When gravitational effects are considered, the virtual particles will attract each other changing their configuration, making it denser, and thus, we expect that gravitational corrections will make the renormalized electrical charge weaker than in usual QED. In mathematical terms, this means that the gravitational corrections will contribute with a term that has a negative sign to the renormalized charge and therefore with a positive term to the beta function.

\begin{figure}[h]
	\begin{center}
	\includegraphics[angle=0 ,width=12cm]{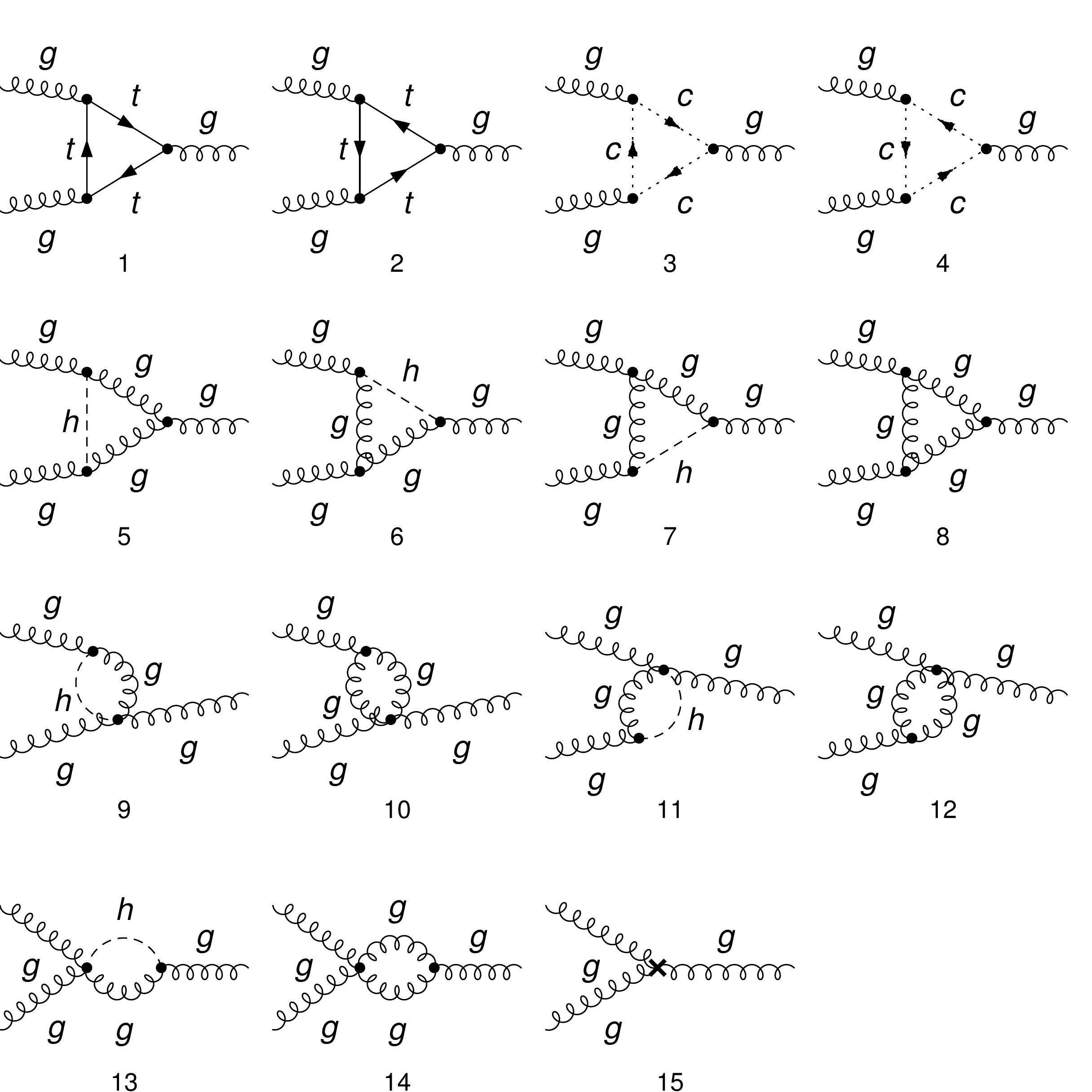}
	\caption{Feynman diagrams to the gluons vertex interaction at one-loop order.}
	\label{fig06}
	\end{center}
\end{figure}

\begin{figure}[h!]
	\begin{center}
	\includegraphics[angle=0 ,width=14.5cm]{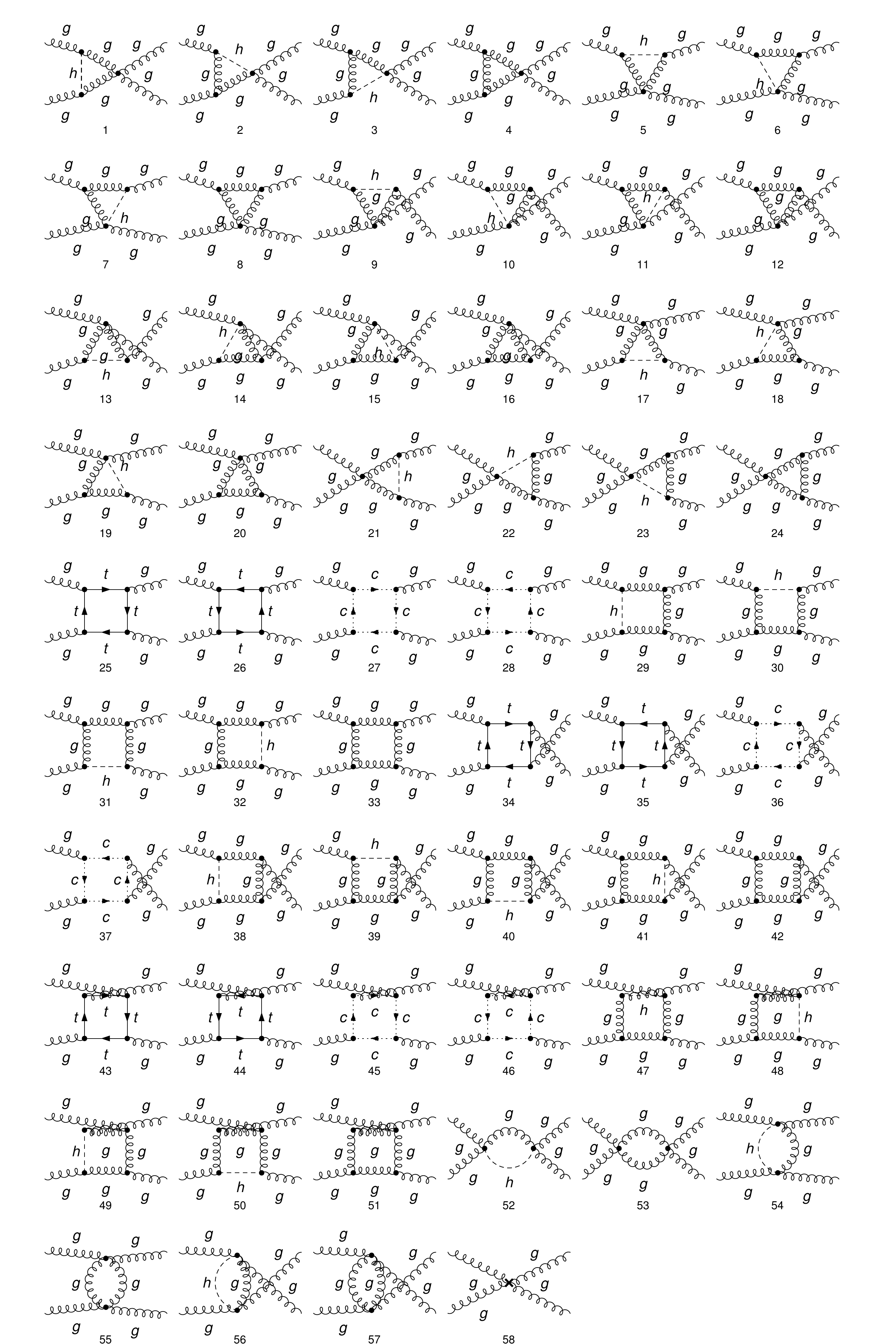}
	\caption{Feynman diagrams to the scattering between gluons up to one-loop order.}
	\label{fig07}
	\end{center}
\end{figure}

\begin{figure}[h!]
	\centering
	\includegraphics[angle=0 ,width=16cm]{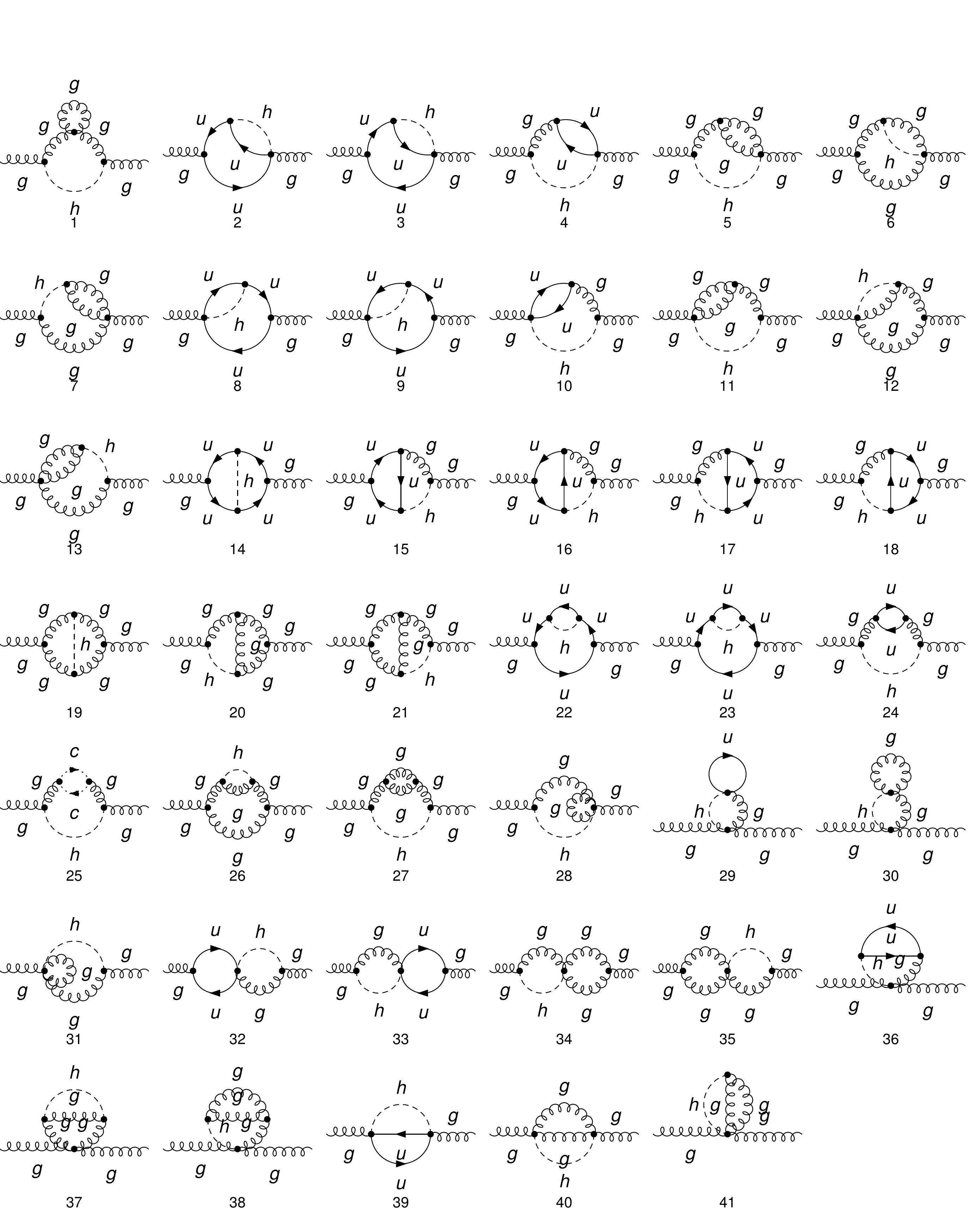}
	\caption{Feynman diagrams to the gluon self-energy involving gravitons at two-loop order.}
	\label{fig08}
\end{figure}

However, when only one-loop corrections are considered, this gravitational effect is not observed in the beta function. The reason is that the polarized particles that make the renormalized charge weaker are virtual particles, then we expect that gravity will change their configuration only when we consider the exchange of gravitons between them and---as we can see diagrammatically from Figs. \ref{fig01}, \ref{fig02}, \ref{fig03}, \ref{fig04}, \ref{fig05}, \ref{fig06} and \ref{fig07}---there is no exchange of gravitons between virtual particles at one-loop order. The situation is different at two-loops, where such exchanges are present (see Fig. \ref{fig08}), resulting in a negative (positive) contribution to the renormalized charge (beta function), as observed in \cite{Bevilaqua:2021uzk}.

Since gravity is always attractive, we expect that the same analysis should be valid for the non-Abelian case. So, although in the present work we have found a vanishing contribution to the beta function due to gravitational effects, we expect that at two-loops order gravitational interaction will produce a positive term to the beta function. This could potentially spoil asymptotic freedom, but this contribution is expected to be very small, as we can see from the result obtained for the QED case \cite{Bevilaqua:2021uzk},
\begin{equation}
 \beta(e) = \frac{e^3}{12\pi^2} + \frac{e^5}{128\pi^4} + \frac{5e^3m^2}{24\pi M_P^2},
\end{equation}
in which we used $\kappa^2 = 32\pi/M_P^2$ where $M_P$ is the Planck mass. Therefore, the asymptotic freedom would be spoiled only for Standard Model extensions in which there are fermions with masses of the order of Planck mass to make the gravitational contribution relevant.

\section{Concluding remarks}\label{sec4}

Now we end the paper by drawing our final considerations and discussing the validity of our results by reviewing discussions from the literature. Our main conclusion from the present work is that the Slavnov-Taylor identities are respected at one-loop order (cf. Sec. \ref{sec3}). This indicates that the color charge is universal: \textit{i.e.}, different processes have the same coupling constant, and therefore, the gauge invariance is preserved.

Let us first consider the operator mixing, \textit{i.e}, the mixing of the contributions from the original coupling with the contribution from a higher order term introduced in the effective Lagrangian to deal with the divergencies. This mixing makes the renormalized coupling to be different if different processes are considered and, therefore, nonuniversal. As it was shown in \cite{Anber:2010uj}, this mixing occurs for the Yukawa coupling ($y\bar{\psi}\psi\phi$) and the authors suggested that the same would occur in other theories, including gauge theories. As shown in Ref. \cite{Bevilaqua:2015hma}, this is indeed the case for the self-interaction coupling constant in the scalar QED. However, our calculation shows that it is possible to separate the contributions from different terms involved in the renormalization of the electric charge because each counterterm has its own kinematics. Thus, we avoid the operator mixing, rendering a universality to the renormalized electric charge. The universality of the gauge coupling was further discussed in \cite{Bevilaqua:2021uev}, in which our results were extended to the gauge coupling for fermionic QED, where again each counterterm has its own kinematics. This indicates that the gauge coupling constants are protected from the mixing by their gauge invariance. In fact, when computing the renormalized electrical charge, we need only to compute the photon self-energy counterterm (due to gauge invariance, we have that $Z_2=Z_1$, and we end up with $e_r = \mu^{-2\epsilon}Z_3^{-1/2}e_0$). Although in the non-Abelian case, one must argue that $Z_1\neq Z_2$ and we thus must consider the operator mixing. To discuss the kinematics of the counterterms more precisely, we intend to study scatterings in this non-Abelian model in a future work.

Gravitational contributions to the running of gauge coupling in the Einstein-Yang-Mills theory were also calculated in Refs. \cite{Tang:2008ah, Tang:2011gz}, where the authors compared different regularization schemes and found that, while dimensional regularization leads to no gravitational contribution at one-loop, the use of loop regularization leads to a nonzero contribution that is proportional to $\mu^2$. They claim that, although the Slavnov-Taylor identities are satisfied irrespective of the regularization scheme used, the gravitational correction for the beta function is scheme dependent. Moreover, they argue that loop regularization is required to properly deal with gravitational interaction because this method is sensitive to the quadratic divergences present in the Feynman diagrams, while dimensional regularization can only treat logarithmic divergences. However, such dependence on the UV cutoff is unphysical and should not be considered in the evaluation of the running coupling if we want to discuss physical processes at energies around that scale \cite{Donoghue:2019clr}. Indeed, as it was pointed out in \cite{Anber:2010uj}, if one considers the S-matrix, the quadratic divergences play no role. Actually, even if one uses the background field method to study quantum gravity, it is possible to define the electrical charge in a physically motivated way that it will not depend on such quadratic divergences \cite{Toms:2011zza}.

The choice of the regularization scheme should not matter for the physical conclusions we draw from the theory, but it does affect the way the renormalized coupling constant is defined. While we used the logarithmic running, usual in quantum field theory, the authors of \cite{Robinson:2005fj, Tang:2008ah, Tang:2011gz} and the community of the asymptotic safety (AS) methods use a power-law running \cite{Bonanno:2020bil, Christiansen:2017cxa}. However, the question whether this kind of running is in fact physical or not is far from trivial. It has already been shown that theories without gravity characterized by a nontrivial fixed point, therefore asymptotically safe, can be well defined using a power-law running \cite{Bond:2017lnq, Bond:2019npq}. Then, one can obtain from them some known results from the standard model as shown in \cite{Hiller:2019mou}, in which the authors used an asymptotically safe extension of the standard model to explain the anomalous magnetic moments from the electron and the muon. However, as pointed out by Donoghue in his critique to the AS program \cite{Donoghue:2019clr}, a power law running is not well defined in a Minkowski space for kinematical reasons. Power law makes different channels go in opposite directions. For example, if one computes the renormalized charges via $s$-channel processes, the results are different from the ones computed from $t$-channel processes. Therefore, one can only define a power law running for the coupling constants in a Euclidian space (as it is done in AS) and the continuation to a Lorentzian signature is challenging for several technical reasons related to the nature of the gravitational interaction \cite{Bonanno:2020bil}. On the other hand, in a recent work~\cite{Fehre:2021eob}, Fehre \textit{et.al.} used a novel Lorentzian renormalization group approach and were able to define an asymptotically safe theory of gravity in a Lorentzian signature using a power law running.
Finally, we would like to discuss the possible implications of considering a cosmological constant. It was shown by Toms \cite{Toms:2008dq,Toms:2010vy} that a positive cosmological constant could generate a negative gravitational contribution to the beta function for the electric charge. But, with the running being only logarithmic, such contribution is phenomenologically irrelevant due to the order of magnitude of observed cosmological constant~\cite{TomsReplies}. Also, a study based on scattering processes was done in \cite{Lehum:2013oja}; the results suggest a negative gravitational contribution to the beta function of the $\lambda$ coupling constant in a $\lambda\phi^4$ model coupled to gravity in the EFT framework. A possible problem in \cite{Lehum:2013oja} is that the calculations were done in the Feynman gauge. Therefore, to fully understand the implications of introducing a cosmological constant in our model, we should compute the S-matrix using an arbitrary gauge. It is our intention to do so in a future work.

\acknowledgments
Huan Souza is thankful to Danilo T. Alves for rich discussions about the physical meaning of renormalization in quantum field theory. The authors would like to thank Sergey Odintsov and Daniel Litim for pointing out several interesting references. The work of HS is partially supported by Coordena\c{c}\~ao de Aperfei\c{c}oamento de Pessoal de N\'ivel Superior (CAPES).

\newpage

\end{document}